\documentclass[longauth]{aa}
\usepackage{txfonts,natbib,graphicx}
\bibpunct{(}{)}{;}{a}{}{,}
\begin{document}
 
\title{The interface between the stellar wind and interstellar
medium around R Cassiopeiae revealed by far-infrared
imaging\thanks{Based in part on observations with 
{\sl AKARI}, a JAXA project with the participation of ESA, and with the
{\sl Spitzer Space Telescope}, which is operated by the Jet Propulsion 
Laboratory, California Institute of Technology under a contract with NASA.}}

\author{%
Toshiya Ueta\inst{1}
\and Robert E.\ Stencel\inst{1}
\and Issei Yamamura\inst{2}
\and Kathleen M.\ Geise\inst{1}
\and Agata Karska\inst{3,4}
\and Hideyuki Izumiura\inst{5}
\and Yoshikazu Nakada\inst{6}
\and Mikako Matsuura\inst{7,8}
\and Yoshifusa Ita\inst{2,9}
\and Toshihiko Tanab\'{e}\inst{6}
\and Hinako Fukushi\inst{6}
\and Noriyuki Matsunaga\inst{6}
\and Hiroyuki Mito\inst{10}
\and Angela K.\ Speck\inst{11}}

\institute{%
Department of Physics and Astronomy, University of Denver, 
2112 E.\ Wesley Ave., Denver, CO 80208, USA
\and Institute of Space and Aeronautical Science, 
Japan Aerospace Exploration Agency, 3-1-1 Yoshinodai, Sagamihara,
Kanagawa 229-8510, Japan
\and Max-Planck-Institut f\"{u}r Extraterrestrische Physik,
Giessenbachstraat 1, 85748 Garching, Germany
\and Leiden Observatory, Leiden University, PO Box 9513, 
2300 RA Leiden, The Netherlands
\and Okayama Astrophysical Observatory, National
Astronomical Observatory, Kamogata, Asakuchi, Okayama 719-0232, Japan 
\and Institute of Astronomy, School of Science, 
University of Tokyo, 2-21-1 Osawa, Mitaka, Tokyo 181-0015, Japan 
\and Department of Physics and Astronomy, University
College London, Gower Street, London WC1E 6BT, UK 
\and Mullard Space Science Laboratory, University
College London, Holmbury St.\ Mary, Dorking, Surrey RH5 6NT, UK 
\and National Astronomical Observatory of Japan, 2-21-1
Osawa, Mitaka, Tokyo 181-8588, Japan 
\and Kiso Observatory, Institute of Astronomy, School of
Science, University of Tokyo, Mitake, Kiso, Nagano, 397-0101, Japan
\and Department of Physics \& Astronomy, University of Missouri, 
Columbia, MO 65211, USA}

\date{Received 14 October 2009/ Accepted 12 November 2009}

\abstract
{}
{The circumstellar dust shells of intermediate initial-mass ($\sim 1$ to
8 M$_{\odot}$) evolved stars are generated by copious mass loss during
the asymptotic giant branch phase.
The density structure of their circumstellar shell is the direct
evidence of mass loss processes, from which we can investigate the
nature of mass loss.} 
{We used the {\sl AKARI Infrared Astronomy Satellite} and the {\sl
Spitzer Space Telescope} to obtain the surface brightness maps of an
evolved star R Cas at far-infrared wavelengths, since the temperature of
dust decreases as the distance from the star increases and one needs to 
probe dust at lower temperatures, i.e., at longer wavelengths.
The observed shell structure and the star's known proper motion
suggest that the structure represents the interface regions between the
dusty wind and the interstellar medium. 
The deconvolved structures are fitted with the analytic bow shock
structure to determine the inclination angle of the bow shock cone.}  
{Our data show that (1) the bow shock cone of $1 - 5\times 10^{-5}$ 
M$_{\odot}$ dust mass is inclined at $68^{\circ}$ with respect to the
plane of the sky, and (2) the dust temperature in the bow shock cone is
raised to more than 20 K by collisional shock interaction in addition to
the ambient interstellar radiation field.
By comparison between the apex vector of the bow shock and space motion
vector of the star we infer that there is a flow of interstellar medium
local to R Cas whose flow velocity is at least 55.6 km s$^{-1}$,
consistent with an environment conducive to dust heating by shock
interactions.}
{}

\keywords{%
Circumstellar matter --- 
Infrared: stars ---
Stars: AGB and post-AGB ---
Stars: individual (R Cas) ---
Stars: mass loss ---
ISM: kinematics and dynamics} 

\authorrunning{Ueta et al.}
\titlerunning{The Wind-ISM Interface around R Cassiopeiae seen in the Far-IR}

\maketitle

\section{Introduction}

Numerous observations have elucidated the magnitude and ubiquity
of mass loss across the upper-right side of the Hertzsprung-Russell
diagram in the fifty years since \citet{d56} discussed the existence of 
blue-shifted circumstellar cores in the spectrum of the red supergiant
star $\alpha$ Her, which constituted one of the first pieces of direct
evidence for high rates of mass loss.
Since such high rates of mass loss among asymptotic giant branch (AGB)
stars rival evolutionary timescales and substantially affect stellar 
evolutionary tracks, careful investigations into mass loss from these
stars are indeed necessary.
Moreover, since mass loss defines the boundary conditions for stellar
evolutionary tracks, the rate of mass loss from these stars is hardly
time-invariant.  
These facts have had theorists perplexed, who are struggling with the
basic challenge of how to lift so much mass away from the gravitational
hold of the star \citep[e.g.][]{gh04}.  

Observations made with the {\sl Infrared Astronomical Satellite} ({\sl  
IRAS}\/) during the 1980's in the far-infrared (far-IR) have
demonstrated that extended shells of evolved stars  -- the anticipated
effect of continuous dusty mass loss -- were present
\citep[e.g.][]{stencel88,y93a}.     
During the 1990s, the {\sl Infrared Space Observatory} ({\sl ISO}) and
ground-based IR work began to refine those results
\cite[e.g.][]{izu97,meixner99}, indicating variations in the mass loss
rate over time that resulted in multiple shells and axisymmetric
structures.    
This decade, we are fortunate to have new instruments with higher
resolution and sensitivity, such as the {\sl AKARI Infrared Astronomy
Satellite} \cite[{\sl AKARI}, formerly known as {\sl
ASTRO-F};][]{murakami07} and the {\sl Spitzer Space Telescope}
\citep[{\sl Spitzer};][]{w04} that can more carefully map out the mass
loss history of evolved stars.   

In parallel with investigations into the mass loss history of evolved
stars, evidence of interactions between the circumstellar matter and
interstellar medium (ISM) around AGB stars is growing, with new
observations of bow shocks around R Hya 
\citep{ueta06} and Mira \citep{martin07,ueta08}, plus theoretical
considerations of the phenomena \citep[e.g.][]{villaver03,wareing07}.  
This report is one of the first of several studies of well-resolved
extended circumstellar dust shells of AGB stars under {\sl AKARI} and
{\sl Spitzer} observing programs labeled ``Excavating Mass Loss History
in Extended Dust Shells of Evolved Stars'' (MLHES).  
In this paper, we explore the detection of an arcminute-sized dust
shell around the oxygen-rich AGB star, R Cassiopeiae (HD 224490;
hereafter R Cas), in context of interactions between the AGB wind from R
Cas and the ISM.

\section{R Cas: the Star and its Circumstellar Dust Shell}
 
The Mira type variable, R Cas, is an oxygen-rich, M7IIIe star with a
period of 431 days \citep{k69} and estimated mass loss rate of $5 \times 
10^{-7}$ M$_{\odot}$ yr$^{-1}$ \citep{km85}.  
This star is known to show an extended circumstellar shell originally
detected by {\sl IRAS} at 60$\mu$m, having angular extent up to
$4\farcm3$ \citep{y93a}.
\citet{bs94} later reported the angular size of $2\farcm8$ using
deconvolution of the point-spread-function (PSF).
The linear extent of the shell depends on the distance determination.  

{\sl Hipparcos} \citep{perryman97} measured a parallax for R Cas to be
$9.37 \pm 1.10$ milli-arcseconds (mas), corresponding to $107 \pm 13$
pc.  
A new calculation done by \citet{vl07} yielded $7.95 \pm 1.02$ mas,
corresponding to $127 \pm 16$ pc.
Other authors prefer values as small as 100 pc
\citep{pourbaix03,knapp03} or as large as 220 pc \citep{km85}.  
{\sl Hipparcos} also detected a 10 mas shift of the centroid of the
star, whose time-dependent, asymmetric size was determined
interferometrically to be 20-40 mas diameter \citep{hofmann01}.
Since the rotation of a bright stellar spot, for example, could increase
the apparent parallax value, the {\sl Hipparcos} measurement for the
distance to R Cas is more uncertain than the quoted error values above.   
\citet{v03} performed VLBI astrometry and obtained a significantly
smaller parallax value ($5.67 \pm 1.95$ mas) for R Cas with respect 
to the {\sl Hipparcos} measurement.
Their analysis suggests the distance of $176^{-92}_{+45}$ pc.
This maser astrometry value is more in line with the larger distance
estimate of 160 pc inferred from the Period-Luminosity relations 
\citep{haniff95}.
Since an ensemble of VLBI measurements over a long time-basis is
less affected by uncertainties induced by changes on smaller time scales
that probably affected {\sl Hipparcos} measurements, we will adopt the
VLBI distance measurement of $176$ pc for our purposes.  

\section{Far-IR Observations of R Cas}

\subsection{AKARI FIS Imaging}

We observed R Cas in the four bands at 65, 90, 140 and $160\mu$m with
the Far-Infrared Surveyor \citep[FIS;][]{kawada07} on-board {\sl AKARI} on
2007 January 16 as part of the {\sl AKARI\/}-MLHES Mission Program (PI:
I.\ Yamamura).   
Observations were made with the FIS01 (compact source photometry) slow-scan
mode, in which two linear strips of forward and backward scans were performed
with a 70$\arcsec$ spacing between the scan strips at the $15\arcsec$
s$^{-1}$ scan rate with a reset rate of 0.5 sec, resulting in the sky  
coverage of roughly $10\arcmin \times 50\arcmin$ centered at the target.  

The FIS Slow-Scan Toolkit \citep[ver.\ 20070914\footnote{Available
at http://www.ir.isas.jaxa.jp/AKARI/Observation/};][]{verdugo07} was used 
to reduce the data. 
We found that the quality of the resulting map was improved when we used
a temporal median filter with the width of 200 s (or longer), a temporal
boxcar filter with the width of 90 s, and the sigma clipping threshold
of 1.5. 
For the reduction of data in the short wavelength bands (at 65 and
$90\mu$m), the results were also improved when we performed
flat-fielding of the data using the local ``blank'' sky data. 

The resulting maps are in $15\arcsec$ and $30\arcsec$ pixel$^{-1}$
(nominal scale) for the short wavelength (65 and $90\mu$m; SW) and
long wavelength (140 and $160\mu$m; LW) bands, respectively, with the
average of 6, 9, 15, and 10 sky coverages per pixel at 65, 90, 140
and $160\mu$m, respectively. 
The resulting 1 $\sigma$ sensitivities and the average
sky emission (the component removed by median filtering during the  
reduction) are found to be 1.2, 0.6, 1.3 and 1.3 MJy sr$^{-1}$ 
and $8.7\pm0.3$, $8.6\pm0.1$, $13.2\pm0.3$ and $9.9\pm0.3$
MJy sr$^{-1}$ at 65, 90, 140 and $160\mu$m, respectively.
Characteristics of the {\sl AKARI} observations and images are
summarized in Table \ref{obschara}: images themselves are shown in
the top row of Figure \ref{imgakari}. 

\subsection{Spitzer MIPS Imaging} 

R Cas observations at 70$\mu$m by {\sl Spitzer} was made with the
Multiband Imaging Photometer for {\sl Spitzer} \citep[MIPS;][]{rieke04}
on 2008 February 18 as part of the {\sl Spitzer}-MLHES project (PI: T.\
Ueta).  
Observations were done in the photometry/fixed-cluster-offset mode, in 
which a series of 12 exposures were made in a spiral pattern around R
Cas.
This exposure pattern was intended not to allow the bright central star
fall on the 70$\mu$m Ge:Ga detector array in ``prime'' and on the
24$\mu$m Si:As detector array in ``non-prime'' in order to avoid
saturation and severe transient effects due to the central star.
In effect, we achieved the sky coverage of $13\arcmin \times
22\arcmin$ centered at the target (but the central $3\arcmin \times
1.8\arcmin$ around the star is unobserved).  

For the data reduction, we started with the basic calibrated data (BCD),
which are relatively free from instrumental artifacts.
However, to optimize the detection of intrinsically faint extended shells
we used a custom IDL script to remove time- and pixel-dependent
instrumental effects still remaining in the BCD. 
This script was developed originally to reduce similar MIPS data
obtained in the MIPS IR Imaging of AGB Dust shells (MIRIAD) project (PI:
A.\ K.\ Speck), based on the idea that it is highly unlikely that there
arises any periodicity in the time-series pixel readings given the way
the {\sl Spitzer\/} aperture was dithered around in the target region. 
This extra BCD cleaning has proven to be effective in correcting those
residual instrumental effects and removing residual background sky
emission \citep[cf.][]{ueta06}.
We then used the MOsaicker and Point source EXtractor (MOPEX) software
\citep[ver.\ 20080819\footnote{Available at
http://ssc.spitzer.caltech.edu/postbcd/};][]{makovoz06} to produce 
a final mosaicked image.  

The resulting map is in 4\farcs92 pixel$^{-1}$ (sub-pixelized by a
factor of 2 from the nominal scale) with the average 
of 9, and up to 24, sky coverages per pixel.
The 1 $\sigma$ sensitivity and the average sky emission (the component
removed during the reduction) are found to be 
1.1 MJy sr$^{-1}$ and $10.0 \pm 0.7$ MJy sr$^{-1}$, respectively.
Characteristics of the {\sl Spitzer} observations and image are
summarized in Table \ref{obschara}: image itself is shown in the
top-left frame of Figure \ref{imgspitzer}.  

\subsection{Photometry and Deconvolution}

Photometry was done only with the AKARI images that capture the
circumstellar shell for its entirety.
Because the absolute calibration of the FIS Slow-Scan data is based on
the measurements of the diffuse sky emission from zodiacal light and
interstellar cirrus done by CODE/DIRBE measurements \citep{verdugo07},
the surface brightness of the extended shell has already been
calibrated. 
However, for the emission core that is essentially a point source a
series of corrections (aperture, flux, and color corrections) needs to
be applied.
Thus, we separated the structure into two parts -- the extended shell
and the emission core -- and performed photometry separately.
For the emission core, we followed a method of aperture photometry
elucidated by \citet{shi09} and evaluated flux correction factors. 
For the extended shell, we simply integrated the surface brightness over
the shell.
However, a care was taken not to double-count the flux component in the 
extended shell part that is accounted for as part of the core emission
via aperture correction.
For each flux value obtained for the core and shell, we independently
applied color correction and obtained the final flux values, which are
listed in Table \ref{obschara}. 

For PSF calibration purposes, we also observed an M5 III giant
$\beta$ Gru \citep{engelke06} with {\sl AKARI} on 2006 November 16  
and 
an asteroid Ceres \citep{ml02} with {\sl Spitzer} on 2008 February 17.
The same instrumental/mapping set-up was used for both PSF observations,
except for {\sl Spitzer} observations, for which the
photometry/moving-cluster-offset mode was used because Ceres is a moving 
target. 
With the observed PSF maps, deconvolution was performed using an
IRAF\footnote{IRAF is distributed by the National Optical Astronomy
Observatories, which are operated by the Association of Universities for
Research in Astronomy, Inc., under cooperative agreement with the
National Science Foundation.} task {\sl lucy}, which is based on the
Lucy-Richardson algorithm. 
The average gain and read-out noise were computed for each map given the 
documented detector responsivity for FIS and MIPS, integration time, and
the averaged number of sky-coverage per pixel. 
Corresponding deconvolved images are shown in the bottom row of Figure
\ref{imgakari} and in the bottom-left frame of Figure \ref{imgspitzer}. 

\section{Results: the Extended Dust Shell of R Cas}

\begin{figure*}
 \centering
  \includegraphics[width=17cm]{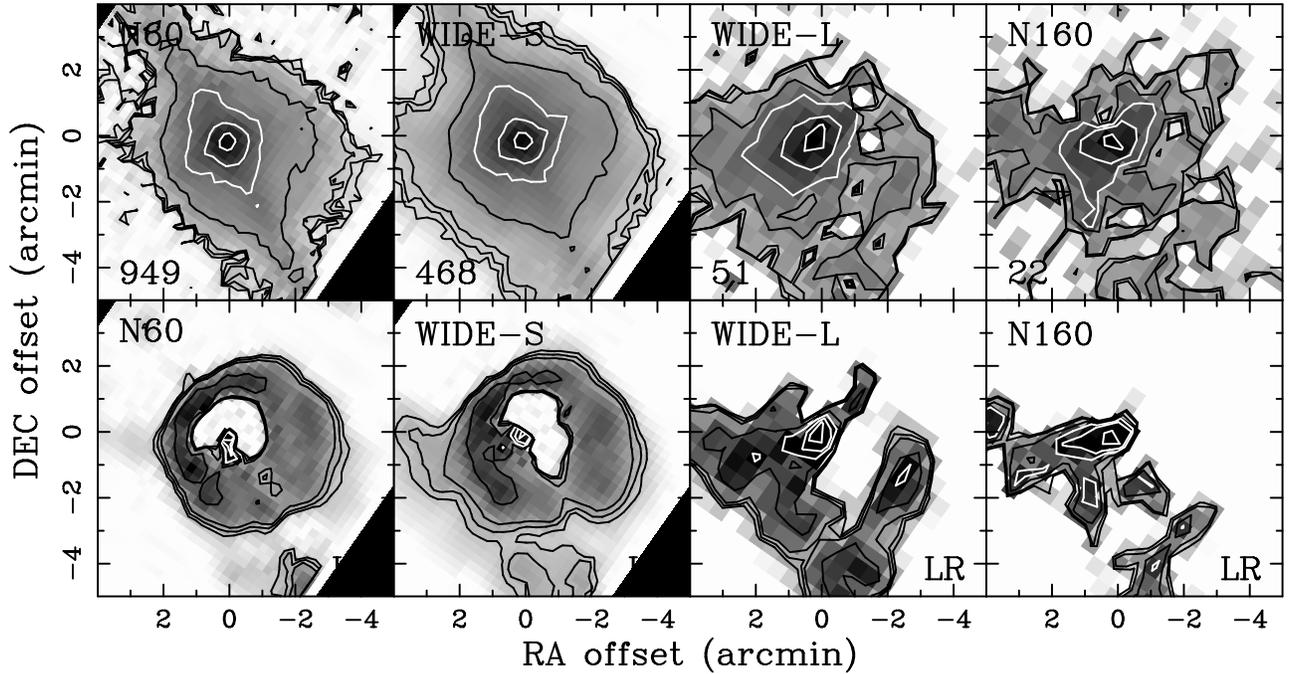}
\caption{AKARI/FIS maps of R Cas in the SW bands -- N60
 (65$\mu$m) and WIDE-S (90$\mu$m) at 15\arcsec pixel$^{-1}$ 
scale -- and in the LW bands -- WIDE-L (140$\mu$m) and N160
(160$\mu$m) at 30\arcsec pixel$^{-1}$ scale -- in the top row from left
 to right, respectively.
Background emission has been subtracted by a combination of
temporal filters during data reduction. RA and DEC offsets
(with respect to the stellar peak) are given in arcminutes. 
The number at the bottom left indicates the peak surface brightness in
 MJy sr$^{-1}$. 
The log-scaled grayscale surface brightness is contoured at 90, 70, 50
 (in white), 30, 10, 3, and 1$\sigma_{\rm sky}$.
North is up, and east to the left.
Images in the bottom row show deconvolved maps at corresponding
 wavelengths, for which ``LR'' on the bottom right indicates the
 Lucy-Richardson algorithm used. 
\label{imgakari}}
\end{figure*}

\begin{figure}
 \centering
  \resizebox{\hsize}{!}{\includegraphics{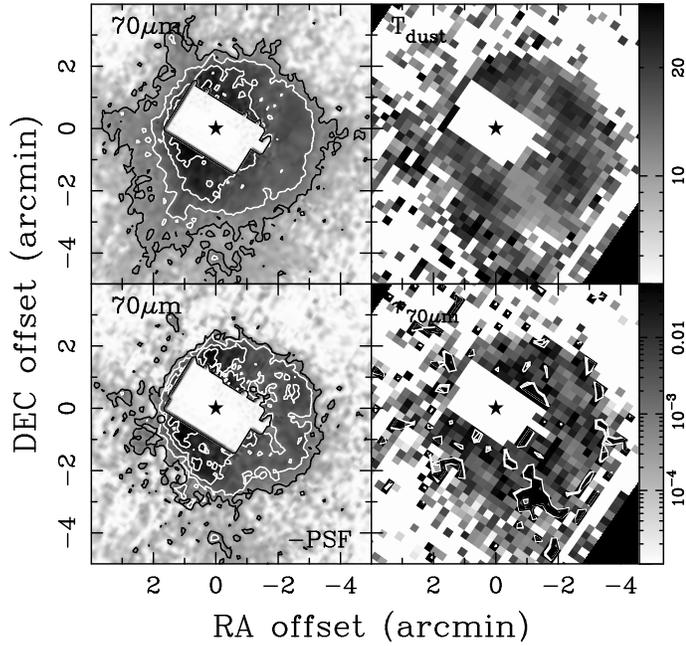}}
\caption{Spitzer/MIPS map of R Cas in the 70$\mu$m
band at 4\arcsec92 pixel$^{-1}$ scale (top left frame) and its
 PSF-subtracted image (bottom left frame).  
Background emission has been subtracted by a combination of
temporal filters during data reduction. RA and DEC offsets
(with respect to the position of the star, indicated by the ``star''
 symbol) are given in arcseconds.  
Also shown are the dust temperature (top right) and optical depth (at
 70$\mu$m; bottom right) maps derived from the fitting of the shortest 3
 waveband images (65, 70, and  90$\mu$m) under the assumption of
 optically thin dust emission. 
The wedges on the right indicate the log-scaled dust temperature in K (top)
and optical depth at 70$\mu$m (bottom).
North is up, and east to the left.  
\label{imgspitzer}}
\end{figure}

Both {\sl AKARI} (Figure \ref{imgakari}) and {\sl Spitzer} (Figure
\ref{imgspitzer}) images of R Cas look very much extended, while the
faint surface brightness in the shell becomes progressively harder to
make out in the {\sl AKARI\/} LW bands, especially at 160$\mu$m.  
The radial surface brightness profile centered at the position of the
star yielded $3\arcmin$ to $4\arcmin$ radius for the {\sl AKARI} SW
bands and {\sl Spitzer} 70$\mu$m band and about $2\arcmin$ radius for
the {\sl AKARI} LW bands (at 3$\sigma_{\rm sky}$; Table \ref{obschara}).  
While the position of the star is obviously off-centered, the extended
shell is roughly circular whose $3\sigma_{\rm sky}$ radius is 2\farcm3
to 2\farcm8 at 65, 70, and 90$\mu$m.
To clarify we also note that the {\sl AKARI} SW band images (N60
and WIDE-S) are affected by the cross-talk of the FIS detector that
manifests itself as a linearly extended emission structure emanating from
the central source into the position angles at $48^{\circ}$ and
$228^{\circ}$ E of N \citep{shi09}.

The shell's emission structure consists of the relatively flat
``plateau'' region on the west side (of surface brightness $\sim 15$ to 20 MJy
sr$^{-1}$ at 65 to 90$\mu$m and $< 10$ MJy sr$^{-1}$ at 140$\mu$m) and
the region of higher surface brightness (emission core) on the east side
around the central star. 
This particular emission structure, however, does not appear to be
caused by the off-centered central star.
The deconvolved maps in the {\sl AKARI} SW bands (the bottom-left images
in Figure \ref{imgakari}) and in the PSF-subtracted {\sl Spitzer}
70$\mu$m band (the bottom-left image in Figure \ref{imgspitzer}) all
show consistently that the surface brightness is enhanced along a
relatively well-defined rim that goes around the periphery of the shell
for almost half the shell on its east side.

Assuming that the detected far-IR emission is mostly of dust continuum,
the surface brightness enhancement on the east side can be caused by
either the density or temperature enhancement of dust grains in the
shell (or both). 
This is because in such an optically thin environment at far-IR
the surface brightness $S_{\nu}$ is proportional to $\tau_{\nu}
B_{\nu}(T)$, where $\tau_{\nu}$ is optical depth along the line of sight
and $B_{\nu}(T)$ is the blackbody function for dust grains at the
temperature $T$. 
Thus, we fit the $\tau_{\nu} B_{\nu}(T)$ curve with the measured surface
brightnesses at 65, 70, and 90$\mu$m to infer the dust temperature
$T_{\rm dust}$ and the optical depth at 70$\mu$m, ${\tau_{70\mu{\rm
m}}}$ simultaneously at each pixel.
This was done by rescaling the {\sl Spitzer} 70$\mu$m map to the same
pixel scale as the {\sl AKARI} SW band images ($15\arcsec$ pixel$^{-1}$)
and performing the fitting using PSF-subtracted maps.
The results are shown in the right frames of Figure \ref{imgspitzer}.

The dust temperature map (the top-right frame of Figure \ref{imgspitzer})
shows the peripheral enhancement similar to the one seen in the
(deconvolved) surface brightness maps (the bottom frames of Figure
\ref{imgakari}).
The dust temperature is the highest ($> 20$K) in the well-defined
peripheral rim on the east side of the shell and in the less
well-defined, clumpier structures on the west side. 
However, the ${\tau_{70\mu{\rm m}}}$ map (the bottom-right frame of
Figure \ref{imgspitzer}) does not show any obvious enhancement near the
periphery of the shell.  
This suggests that the observed surface brightness enhancement in the
extended shell of R Cas is due to the temperature enhancement in the
shell rather than the density enhancement. 
What would cause such a temperature enhancement in the shell, then? 

\citet{v03} measured proper motion of R Cas, $(\mu_{\alpha},
\mu_{\delta}) = (80.52\pm 2.35, 17.10\pm 1.75)$ mas yr$^{-1}$,
by VLBI astrometry.
This translates to the position angle of $78\fdg0 \pm 1\fdg6$ east of
north, which agrees remarkably well with the direction along which
there is a positive gradient of surface brightness.
In other words, the emission structure of the extended shell appears to
show mirror symmetry with respect to a line defined by the direction of
proper motion. 
Interestingly, the direction of the measured proper motion is parallel  
to the direction of the apparent shift of the central star from the
geometric center of the shell.
The shape of the circumstellar shell was fit by an ellipse using 
the MIPS 70$\mu$m map, in which the pixel resolution is the highest.
We defined the shell edge at the 3$\sigma_{\rm sky}$ level and measured
the radius (the distance from the star to the edge) in all directions at
a certain azimuthal interval.
The best-fit semi-major axis length ($a$) and eccentricity ($\epsilon$)
were thus searched for by adjusting these two values while minimizing
the difference between the measured shape and the ellipse defined by the 
particular $(a,\epsilon)$ pair.  
During this fit, we assumed that the central star would fall on one of
the foci and the semi-major axis would lie along the vector defined by
the direction of the measured proper motion. 

The best-fit semi-major axis and eccentricity pair, $(a,\epsilon)$,
turned out to be $a = 165\farcs3$ and $\epsilon = 0.3$ (hence, the
semi-minor axis length $b = 157\farcs7$):
indeed, the extended shell of R Cas is not quite circular in projection.  
According to this best-fit ellipse, the distance from the ellipse center
to one of the foci is $c = a \times \epsilon = 48\farcs8$.
This means that the central star is $48\farcs8$ displaced from the
ellipse center over the course of the shell expansion.  
At the preferred distance for R Cas, 176 pc \citep{v03}, the semi-major
and semi-minor axes correspond to 0.13 to 0.14 pc.
If we assume that the measured CO expansion rate of 12 km s$^{-1}$
\citep{bfo94} is the (constant) expansion velocity of the extended dust
shell, the crossing time of the shell is therefore roughly $10^4$ years.  

If the elongation of the R Cas shell is solely due to the motion of the
central star with respect to the shell (in an otherwise stationary local
environment), the star must have been moving roughly at 5 mas yr$^{-1}$,
which is much less than the observed proper motion of 82.3 mas
yr$^{-1}$.  
Thus, the elliptical elongation of the shell and the offsetting of the
central star do not seem to be self-inflicted, as in the case where
a pile-up of AGB wind material occurs at the interface between
fast and slow AGB winds and defines the edge of the observed shell
\citep{Steffen98}.  
Rather, we speculate that the shell shaping is orchestrated by the 
interactions between the AGB winds and ambient ISM, as in the AGB
wind-ISM interaction discovered around an AGB star, R Hya, by {\sl
Spitzer} \citep{ueta06}. 
In this scenario, the temperature enhancement seen in the east side of
the shell is due to dust heating along the contact discontinuity between 
the AGB wind and the ISM flow.
The temperature enhancement seen in the west side is then probably due
to dust heating in the wake of the AGB wind-ISM interactions flowing
downstream \citep[e.g.][]{wareing07}.
This dust heating is likely due to collisions rather than shock
emission because there is no known H$\alpha$ emission source at the
position of R Cas \citep{fin03}.
This interpretation is consistent with recent \ion{H}{I} observations of
R Cas \citep{mr07}, which identifies a head-and-tail structure whose
alignment in the western wake of our far-IR maps is nearly perfect .

Unlike the R Hya case, however, we do not see any clear parabolic
bow shock structure that is expected to arise at the interface between
the AGB wind and ISM. 
Still, we recognize rather round temperature-enhanced regions toward the
windward direction in the shell and wake-like temperature-enhanced
structures toward the leeward direction.
This is similar to the bow shock of Betelgeuse, for which its bow shock
appears rather circular due to the inclination angle of the shock
surface \citep{u08}.
Assuming momentum conservation across a physically thin (i.e.\
radiative-cooling dominating) shock interface between the AGB wind and
ISM, one can express the bow shock shape analytically as a function of
the latitudinal angle $\theta$ measured from the apex of the bow with
respect to the position of the central star as follows \citep{wilkin96}: 
\begin{eqnarray}\label{bow}
 R(\theta) = R_0 \frac{\sqrt{3(1-\theta\cot\theta})}{\sin\theta} 
\end{eqnarray} 
where $R_0$ is the {\sl stand-off distance} between the star and bow
apex defined as 
\begin{eqnarray}\label{standoff}
 R_0 =\sqrt{\frac{\dot{M}v_{\rm w}}{4\pi\rho_{\rm ISM} v_{*}^2}}
\end{eqnarray}
for which $\dot{M}$ is the rate of mass loss,
$v_{\rm w}$ is the isotropic stellar wind velocity, 
$\rho_{\rm ISM}$ is the ambient ISM mass density, and 
$v_{*}$ is the space velocity of the star.
Because the observed enhancement in the shell (i.e.\ an inclined bow
shock) represents a conic section of the bow shock paraboloid (in
optically thin dust distribution the column density becomes the greatest
at the cross section of the structure with the plane of the sky), we can
determine the inclination angle by fitting the apparent shape of the bow
shock with the analytical function and a given inclination angle
\citep{u08}. 

The best-fit of the Wilkin solution fitting yields the inclination of
the bow shock to be $\theta_{\rm incl} = \pm (68^{\circ}\pm2^{\circ})$
and the position angle of $74^{\circ}\pm2^{\circ}$.
The inclination angle has a degeneracy because this fitting method alone
would not determine if the structure is inclined to us or away from us
with respect to the plane of the sky.
Nevertheless, the direction of the relative motion of the AGB wind shell
with respect to the ambient ISM determined from the apparent bow shock
structure is very much different than inferred from the observed motion
of the central star ($\theta_{\rm incl} = 18\fdg4$ toward us at the
position angle of $78\fdg0$).
The discordance between these two values can be resolved only by
incorporating the idea that the ISM itself is flowing in a particular
direction.
In other words, the difference between the vector of the apparent
(heliocentric) space motion of the central star $\mathbf{v}_{*,\odot}$
and the vector of the space motion of the star {\sl relative} to the
ambient ISM $\mathbf{v}_{*,{\rm ISM}}$ (as derived from the 3-D
orientation of the shock structure) is the heliocentric flow vector of
the ambient ISM $\mathbf{v}_{{\rm ISM},\odot}$:
\begin{eqnarray}\label{vism}
 {\mathbf v}_{{\rm ISM},\odot} = {\mathbf v}_{*,\odot} - {\mathbf
  v}_{*,{\rm ISM}}. 
\end{eqnarray}

The stand-off distance of $1\farcm4\pm0\farcm1$ translates to 0.1pc
at the adopted VLBI-measured distance of 176pc.
Solving eq.(\ref{standoff}) for $v_{*,{\rm ISM}}$ with known quantities 
($\dot{M} = 5 \times 10^{-7}$ M$_{\odot}$ yr$^{-1}$, $v_{\rm w} = 12$ km
s$^{-1}$, and $\rho_{\rm ISM} \approx 1.4 m_{\rm H} n_{\rm ISM}$, where
$1.4 m_{\rm H}$ is the average ISM particle mass and $n_{\rm ISM}$ is
the ISM number density), we have
\begin{eqnarray}
  {\mathbf v}_{{\rm ISM},\odot} 
=
\left(
\begin{array}{r}
v_{{\rm ISM},\odot}^{\rm rad}\\
v_{{\rm ISM},\odot}^{\alpha}\\
v_{{\rm ISM},\odot}^{\delta}
\end{array}
\right)
=
\left(
\begin{array}{r}
-22.9 \\
+67.4 \\ 
+14.3 
\end{array}
\right)
-
\frac{1}{\sqrt{n_{\rm ISM}}}
\left(
\begin{array}{r}
\pm33.0 \\ 
+12.6 \\ 
+3.6 
\end{array}
\right)
~~(\mbox{km s$^{-1}$})
\end{eqnarray}
as a function of the remaining unknown, $n_{\rm ISM}$.
The ${\mathbf v}_{{\rm ISM},\odot}$-$n_{\rm ISM}$ relation suggests that 
$v_{{\rm ISM},\odot}$ reaches its minimum of 55.6 km s$^{-1}$ at $n_{\rm
ISM} = 0.58$ cm$^{-3}$ when the bow shock cone points to us and 72.5
km s$^{-1}$ at $n_{\rm ISM} = 75.5$ cm$^{-3}$ when the bow shock cone
points away from us.
In general, high values are unlikely for both ${\mathbf v}_{{\rm
ISM},\odot}$ and $n_{\rm ISM}$.
Therefore, the case where the bow shock cone points away from us seems
unlikely. 
If this is true the direction of the ISM flow should be going into the
plane of the sky (positive $v_{\rm rad}$), given the space motion of the
central star.
Hence, the ambient ISM number density needs to be less than 2 cm$^{-3}$,
which is in line with the fact that a relatively higher $n_{\rm ISM}$
value is expected for R Cas, a low galactic latitude object.

In the frame of stellar winds at 12 km s$^{-1}$ \citep{knapp98},
the relative velocity of the ambient ISM flow into the interface regions
is $12+35.5/\sqrt{n_{\rm ISM}}$ km s$^{-1}$, which translates into at
least 37 km s$^{-1}$ if $n_{\rm ISM} > 2$ cm$^{-3}$.
Such a velocity could induce relatively weak (collisional) shock
interactions that can raise the dust temperature of the AGB wind-ISM
interface regions to $\sim 20-30$ K as seen from the present far-IR data. 
Given the nature of the central star, it is highly unlikely that dust
heating is solely due to radiation from the central star.
Typically the ambient interstellar radiation field is thought to
contribute in heating dust grains in the outermost regions of the
extended circumstellar shells.
The present analysis of the far-IR data indicates that the AGB wind-ISM
interactions can contribute as yet another source for dust heating in an
otherwise cold environment.
In future radiative transfer studies with our far-IR data sets, we will
quantify relative contributions of these dust heating sources at the
periphery of the extended dust shells around evolved stars as they would
impact the physical and chemical conditions of the AGB wind ejecta upon
injection into the ISM. 

The dust temperature and optical depth maps together with a surface
brightness map would yield crude estimates of dust mass in the shell:
\begin{eqnarray}
 M_{\rm dust} \approx 
  \frac{F_{\nu} \lambda^2 D_{*}^2}{2 k T_{\rm dust} \kappa_{\nu}}
\quad \mbox{or} \quad
\frac{\sum \tau_{\nu,{\rm i}} A_{\rm i}}{\kappa_{\nu}}
\end{eqnarray}
where
$F_{\nu}$ is the measured flux at a given frequency $\nu$ of the band,
$\lambda$ is the wavelength of the band,
$D_{*}$ is the distance to the source,
$k$ is Boltzmann constant,
$\kappa_{\nu}$ is the dust opacity,
$A_{\rm i}$ is the area subtended by each pixel,
and the summation over pixel ${\rm i}$ refers to integration over the shell.
Here, we assume a spherical grain ($0.1\mu$m radius) of olivine
\citep[MgFeSiO$_4$;][]{dor95} and calculated its opacity with Mie theory
(48, 41, and 25 cm$^2$ g$^{-1}$ at 65, 70, and 90$\mu$m, respectively)
and use the adopted distance to R Cas of 176 pc.
Our data then suggest that the dust mass in the shell is roughly $1 - 5
\times 10^{-5}$ M$_{\odot}$.
Given the mass loss rate of $5 \times 10^{-7}$ M$_{\odot}$ yr$^{-1}$ and
the dynamical age of the shell of $\sim 10^4$ years, the rough estimate
of the total mass in the shell is $5 \times 10^{-3}$ M$_{\odot}$,
yielding the gas-to-dust ratio between 100 and 500.
The amount of total mass estimated this way is a lower limit because the 
duration of mass loss could have been more than $10^4$ years.
Then, the estimated gas-to-dust ratio will be correspondingly higher, 
indicating that there exists less dust in a colder environment of the
extended dust shells.
However, this is based on a very rough estimate and needs to be
confirmed by radiative transfer calculation incorporating at least
dust heating by the interstellar radiation field.

\section{Conclusion}

We obtained far-IR images of an oxygen-rich Mira variable R Cas at
65, 70, 90, 140, and 160$\mu$m using {\sl AKARI} and {\sl Spitzer}
and revealed its very extended ($2\arcmin$ to $3\arcmin$ radius,
corresponding to 0.1 pc at its adopted distance of 176 pc), slightly
elliptical ($\epsilon = 0.3$) dust shell, in which the central star is
located offset from the geometric center of the shell in the direction
of the measured proper motion of the central star.
We recognize a positive gradient of the surface brightness along the
direction of the ``shift'' of the central star, which apparently is
caused by the surface brightness enhancement along the periphery of the
shell in the east side seen in deconvolved images.

Fitting of the surface brightness using data in the 3 shortest bands
suggests that the observed enhancement is caused by the temperature
enhancement rather than the density enhancement, prompting a need to
warm up dust grains primarily on the east side of the outer rim. 
Given the coincidence between the direction of the proper motion of the
central star and the direction of the apex of the peripheric surface
brightness/temperature enhancement in the shell, we infer that the
observed shell structure represents the contact surface of the AGB
wind-ISM interaction which is inclined to give an overall spherical
shape instead of a typical parabolic bow shock structure.
This AGB wind-ISM (collisional) interaction therefore warm up dust
grains in the interface regions, causing the temperature enhancement
towards the windward direction of the shell.
Using the maps, we also estimated the total dust mass in the shell to be
$1 - 5 \times 10^{-5}$ M$_{\odot}$.

The shape of the observed enhancement was fitted with the analytical
function for the bow shock cone to derive the inclination angle of
$68^{\circ}$.
The apex vector of the bow shock cone and the space motion vector of
proper motion of the central star were compared to deduce that there is
an ISM flow local to R Cas that has a flow velocity of at least 55.6 km
s$^{-1}$.
Then, the relative velocity of the ambient ISM flow with respect to the
AGB wind-ISM interface regions is at least 37 km s$^{-1}$. 
Therefore, such weak shocks can play a role in heating dust grains in
the outermost regions of these extended dust shells around evolved
stars in addition to the interstellar radiation field that is originally 
expected to play a role in an environment where luminosity from the
central source is not enough for required dust heating.

\begin{acknowledgements} 
We are grateful for financial support from the Institute of
 Space and Astronautical Science under the auspices of the Japan
 Aerospace Exploration Agency 
as well as the Jet Propulsion Laboratory/California Institute of
 Technology.  
\end{acknowledgements}

\clearpage

\begin{table*}
\caption{\label{obschara}Characteristics of the Far-IR Observations of R
 Cas with {\sl AKARI} and {\sl Spitzer}}
\centering
\begin{tabular}{cccccccccc}
\hline\hline
Date \& Time   &
   &
$\lambda$    &  
$\Delta\lambda$    &  
Pixel Scale    &  
Sky Coverage    &  
$S_{\nu, {\rm sky}}$    &  
$\sigma_{\rm sky}$  &
$R_{3\sigma}$ &
$F_{\nu}$ \\  
(UT)    &
Band    &  
($\mu$m)    &  
($\mu$m)    &  
(arcsec)    &  
(pixel$^{-1}$)    &  
(MJy sr$^{-1}$)    &  
(MJy sr$^{-1}$) &
(arcmin) &
(Jy)\\
\hline
\multicolumn{10}{c}{{\sl AKARI\/}/FIS Observations}\\
\hline
2007 Jan 16 10:19 & N60    &  65 & 22 & 15 &  6 &  8.7 & 1.2 & 3.1 & 121.48$\pm$19.90\\
                  & WIDE-S &  90 & 38 & 15 &  9 &  8.6 & 0.6 & 3.8 & 73.42$\pm$10.03\\
                  & WIDE-L & 140 & 52 & 30 & 15 & 13.2 & 1.3 & 2.0 & 25.89$\pm$2.09\\
                  & N160   & 160 & 34 & 30 & 10 &  9.9 & 1.3 & 1.8 & 22.41$\pm$7.47\\
\hline
\multicolumn{10}{c}{{\sl Spitzer\/}/MIPS Observations}\\
\hline
2008 Feb 18 04:02 & MIPS70 &  70 & 19 &  4.92 & 9 & 10.0 & 1.1 & 3.5 & \dots\\
\hline
\end{tabular}
\end{table*}


\clearpage

\object{R Cas}

\end{document}